# An Ethical eValuation Agent (EeVA): Results of a Proof-of-Concept Test on a Prototype Agentic-like Workflow to Assist Ethical Deliberations

**Stephen Milford,[1],[2] B. Zara Malgir,[1] Miguel Vazquez,[3]**

**Abstract –** Ethical deliberation is often misunderstood as a process of producing single 'right' and 'wrong' answers. This creates practical difficulties for non-ethically trained personnel who must respond to ethically laden challenges. We developed EeVA, an agentic-like LLM-based workflow designed to support comparative ethical reflection rather than deliver definitive ethical answers. Methods: EeVA was programmed in n8n using three interconnected workflows: a starter, worker, and emitter workflow. It evaluated uploaded use-cases against 10 ethical frameworks using an evaluator and synthesising prompt. Proof-of-concept testing was conducted on three published use-cases from urban mobility, peer-to-peer energy trading, and social-service resource allocation. Results: Across all three use-cases, EeVA generated consistently structured framework-specific evaluations and integrated syntheses. Outputs differentiated clearly between ethical frameworks, highlighted convergences and divergences, gave recommendations for modification to increase convergences, and noted unresolvable tensions. Syntheses were readable for non-specialists and oriented users away from simplistic 'right' or 'wrong' answers toward reflection on design conditions, safeguards, and areas where full cross-framework alignment was unlikely. Discussion: The findings suggest that LLMs can be organised into usable workflows that preserve ethical plurality while helping bridge the communicative mismatch between ethicists and non-ethically trained personnel. EeVA's value lies not in replacing ethicists or resolving moral disagreement, but in providing a scaffold for structured ethical deliberation. Conclusion: EeVA offers a promising proof of concept for supporting ethical reflections in contexts where access to ethical resources is limited. However, further work is needed on reproducibility, human evaluation, user testing, and efficiency before the system can be considered a mature tool.

## I. INTRODUCTION

Ethical deliberation is often a misunderstood phenomenon. At its core, it is a form of structured reflection on the moral dilemmas we face. For the uninitiated (the non-ethically trained outsider), it appears to be a process used to identify 'right' from 'wrong.' This, however, is not how ethical deliberation functions. Ethics does not comprise a single, universally accepted understanding of truth, and as such cannot provide an objective, definitive answer to morally significant problems. While this point can be applied to diverse fields (e.g., social contexts, political challenges, public policy, etc.), to illustrate our point, this paper will consider the context of technological development.

All technology involves social actors and as such is not ethically neutral (Selbst et al. 2019). Consequently, those who develop technologies must be accountable for the ethical implications of those technologies. Many organisations take this responsibility seriously and implement processes such as internal auditing, documentation, and structured reflections as complements to external governance (Raji et al. 2020). Yet ethical reflection around technological development remains a serious challenge.

Beyond resource constraints and a focus on quantifiable deliverables, an expectational mismatch between ethicists and those who develop new technologies exists. This mismatch leads to serious challenges in cross-field communication. Anyone who has sat in an 'ethics work package' meeting involving trained ethicists and non-ethically trained technical personnel (NTTP) – such as control/automation engineers, programmers, roboticists, designers, etc. – will recognise this mismatch. NTTP ask for determinate conclusions about what is or is not an ethical solution to technical challenges so as to encode this into technical specifications. Ethicists respond with hyper-qualified exceptions, distinctions, or seemingly unrelated definitions about the challenge. The frustration is mutual. NTTP experience ethical input as underspecified, too abstract to quantify, or too dynamic, while ethicists experience technical requests as categorical mistakes misrepresenting the role and function of moral deliberation in society.

The requests of NTTP to have single, or limited sets of, ethical response(s) to questions raised by their technical proposals is not inherently misguided. Technological developments must be operational, specific, and testable. Trade-offs must inevitably be made if we are to progress. However, the expectation that ethics can reliably yield a single, univocal 'right answer' that can be translated into procedural process is a misunderstanding of how moral reasoning functions.

Numerous attempts have been made to deliver ethical solutions to NTTP as single, portable constraint sets. All suffer from serious limitations. At one extreme, ethics is swollen to abstract principles with very little concrete application (fairness, transparency, trust, explainability etc.). At the other extreme, ethics is reduced to a narrow decision rule that focuses on risk but misrepresents the broader moral context. This reduction is encapsulated in the contemporary critiques of 'principle first' design (Mittelstadt 2019; Whittlestone et al. 2019). A number of examples of this challenge could be mentioned, notably attempts to provide algorithmic solutions to the family of trolley-problem dilemmas within autonomous vehicles (AVs). The most famous of these is the Moral Machine experiment (MME; Awad et al. 2018). With millions of responses, the authors were exuberant in their solution, pointing out the inability of ethicists to agree on a solution and their insistence that a system of public voting should be enacted to solve these ethical conundrums (Noothigattu et al. 2018). This approach has been widely criticised (Harris 2020; Etienne 2021; Furey and Hill 2021; Etienne 2022). Today the MME stands as a prominent example of the difficulties and potential pitfalls of attempting to resolve moral dilemmas using computational methods.

[1] Institute for Biomedical Ethics, Basel University, Switzerland; [2] North-West University, South Africa; [3] Barcelona Supercomputing Center
* Corresponding author: Milford.Stephen@gmail.com

The rise of Large Language Models (LLMs), particularly those deployed in readily accessible chatbot AIs such as ChatGPT, Gemini or Claude, may offer a solution. Foundation models such as these can generate summaries, explanations, counter-arguments, and alternative framing at incredible speed, potentially functioning as scaffolds for rapid ethical reflection. Indeed, the field of AI ethicists is blooming. For more than a decade, attempts have been made to build 'artificial moral advisors,' with all the associated meta-ethical concerns (Giubilini and Savulescu 2018; Lara 2021; Lara and Deckers 2020; cf. O'Neill et al. 2023). Today, ethical AI models such as Delphi, and even ethics chatbots fine-tuned by eminent ethicists such as Peter Singer, are easily accessible to the public (Ghose and Singer 2024; Jiang et al. 2025). Work is currently underway to develop benchmarks for the moral reasoning of LLMs (Ji et al. 2025; Jiao et al. 2025). Yet these approaches have received severe critique (O'Neill et al. 2023). Setting aside the well-known challenges of AI hallucinations, Fitzgerald convincingly argues that accuracy cannot be a metric for ethics modelling in LLMs (Fitzgerald 2024). This is because a focus on model accuracy fails to understand the role and function of ethical deliberation.

Ethical dilemmas are closer to what planning theorists call 'wicked' than 'tame' problems (Rittel and Webber 1973). 'Wicked' problems are challenges whose very formulation is contested and whose solutions are not presentable as true/false, or right/wrong. Here, attempted solutions often reshape the problems themselves. The challenge of 'wicked' problems is compounded by the way moral reasoning works. Moral reasoning is not a single method. It can be approached from multiple perspectives depending on presuppositions. In moral philosophy, first principles – those initial (often unconscious) 'self-evident' truths or axioms – form the basis upon which all further moral argumentation is derived. The first principle taken by an ethicist is often a site of dispute with each debating partner arguing for or against the coherence of their respective first principles. From these first principles, different ethical frameworks can be formed that then radically shape responses to ethical dilemmas.

A rudimentary reading of the Stanford Encyclopedia of Philosophy provides a well-founded generalisation as to how these first principles are operationalised (Zalta and Nodelman 2026). For example, utilitarianism operationalises the principle of utility (understood as the promotion of the greatest good/happiness); deontology, the principle of categorical imperative (understood as a universal law); virtue ethics, the principle of the golden mean (a virtuous balance between two extremes); care ethics, the reality of relational ontologies (the inherent interconnectedness of all persons); while contractualism operationalises the first principle of reasonable rejectability (acts are wrong if they violate a set of rules for the general regulation of behaviour that could not be reasonably rejected).

These distinct frameworks are not mere academic anomalies, they are encoded different moral principles that radically impact the solution one may operationalise to address an ethical dilemma. The same intervention will look well-motivated under one framework and morally indefensible under another. Not because the one is superior to the other (although proponents may often argue as such) but because moral reality is multidimensional. This is not a defect in the field of ethics. It is a central part of the epistemic function of ethics. Ethics is often less about decision distillation, and more about a disciplined way of clarifying concepts, surfacing assumptions, identifying stakeholders, articulating options and exposing hidden trade-offs.

One might think of ethics as the air-filtration system in the presidential Situation Room (or more accurately the Presidential Emergency Operations Centre - PEOC). Moral reasoning does not make or recommend decisions, it merely improves the atmosphere in which those decisions are made – what can be said, what must be justified, and which harms are rendered visible or remain hidden. Moral reasoning does not guarantee right decisions, it improves the likelihood of better decisions by shaping deliberation and encouraging rigorous critical thinking.

Consequently, an ethically laden technical challenge faced by NTTP should be evaluated by a range of ethical frameworks on the basis of multiple first principles. Not to provide an absolute right or wrong solution, but to give the development team the opportunity to reflect on their technical proposals so as to increase the chances of better design decisions. A key challenge is that not all operational teams have access to a PEOC staffed with trained ethicists. Ethical expertise and time are scarce resources in many technological contexts. NTTP often cannot sustain continuous embedded ethical consultation, nor treat every design choice as the occasion for a seminar on moral theory.

What is needed is a framework-agnostic tool to assist NTTP to evaluate their operational decisions from the perspective of multiple ethical frameworks and to encourage ethically coherent critical reflection around their particular operational challenge. This paper will outline a proof of concept of such a system. We report on an agentic-like LLM-based workflow prototype to support NTTP in ethical deliberation. It demonstrates the potential for LLMs to address the communicative mismatch described above, and to provide NTTP with a practical tool to evaluate a proposed technical solution (use-case) against a range of ethical frameworks without requiring the end-user to (a) master moral philosophy, or (b) secure sustained input from multiple ethicists for every use-case. The aim of this tool is not to produce a single (or limited set of) right or wrong answers to a use-case, nor to affirm or negate a technical solution proposed by NTTP. Rather it is to help NTTP locate their proposed solutions within a broader ethical landscape and to better reflect on their proposals.

## II. METHODS

**Study purpose and design:** This study aimed to develop and conduct a proof-of-concept test of an LLM-based artefact designed to support cross-framework evaluation and reflection on technical proposals to address ethically laden challenges. This aim followed the broader conceptual position outlined in the introduction, namely that ethical reasoning is multifaceted, interpretive, and comparative rather than reductive and prescriptive.

**Prototype architecture:** The prototype (termed EeVA – the Ethical eValuation Agent) was designed using n8n – an extensible, node-based automation tool that enables the integration of unrelated APIs and computational services into a cohesive workflow. The system architecture comprised three interconnected workflows: *starter*, *worker*, and *emitter*.

The *starter* workflow (see fig. 1) accepted a use-case in the form of a text-based PDF, validated the input, generated a unique run identifier, inserted an initial record into the data table, and returned an immediate response to the user interface indicating the start of the process.

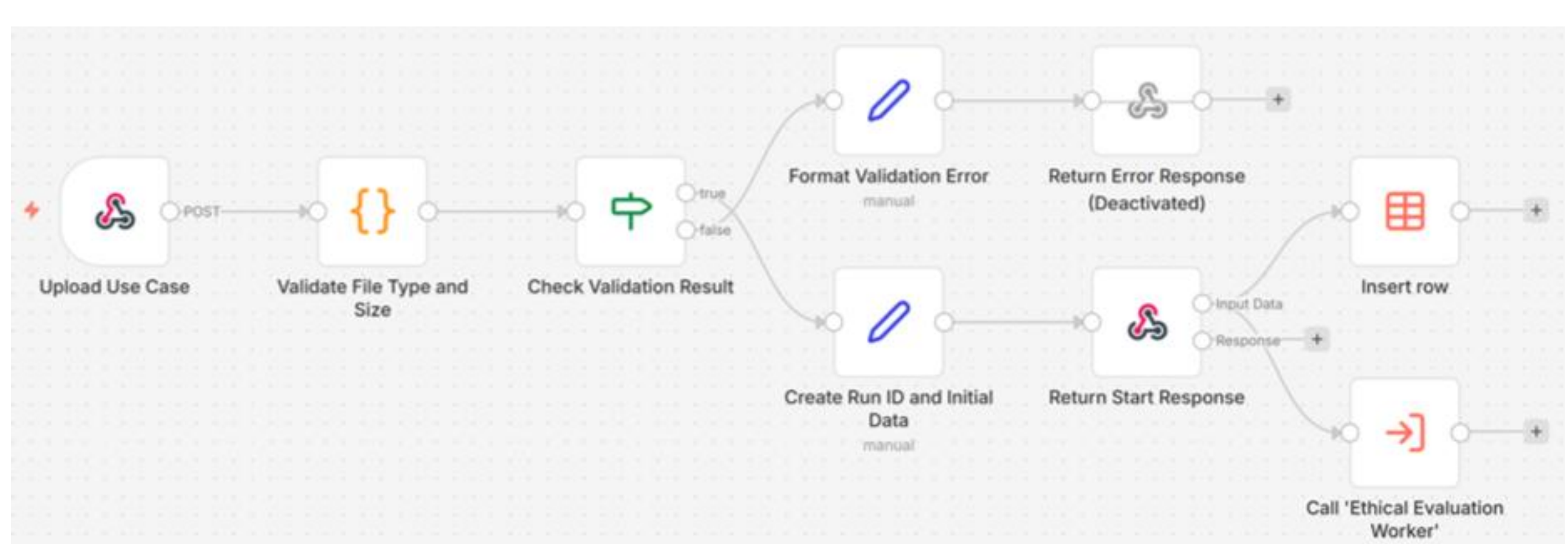

*Figure 1: Starter Workflow*

Having received the file and metadata from the starter workflow, the worker workflow (see fig. 2) processed the input by extracting the source text, generating framework-specific evaluations and synthesis, as well as updating the existing run record. To generate evaluations aligned with specific ethical frameworks, the use-case, the specific framework descriptions, and the prompt were submitted to 10 LLMs, each utilising a distinct ethical framework. Using OpenAI's GPT-4.1 mini model, each LLM generated an evaluation based on its assigned framework. These evaluations were subsequently submitted – together with the original use case and synthesis prompt – to a separate synthesising LLM model (GPT-5.4), which integrated them into a consolidated assessment for non-technically trained end-users.

The *emitter* workflow (see fig. 3) provided a separate retrieval layer, allowing outputs to be fetched by the run identifier without interfering with the upload or processing stage (worker workflow). The separation of upload, processing, and retrieval was deliberately implemented to improve system efficiency and reduce unnecessary execution load in n8n. A simple HTML-based user-interface was designed to upload use-case files, display the use-case name, run ID, synthesis and framework-specific evaluations.

**Framework selection and development:** 10 ethical frameworks that are well-established within moral philosophy were chosen: deontology, utilitarianism, Rawlsian justice, rights-based ethics, contractualism, principlism, virtue ethics, care ethics, communitarianism and casuistry. In line with the project's aim, the intention of framework development was not to reproduce the full complexity of each philosophical tradition but to provide sufficiently robust and recognisable accounts of each framework for comparative evaluation. Therefore, framework descriptions provided to the LLMs were taken from authoritative secondary sources such as the Stanford

Encyclopedia of Philosophy (Zalta and Nodelman 2026) or the Internet Encyclopedia of Philosophy (Fieser and Dowden, n.d.).

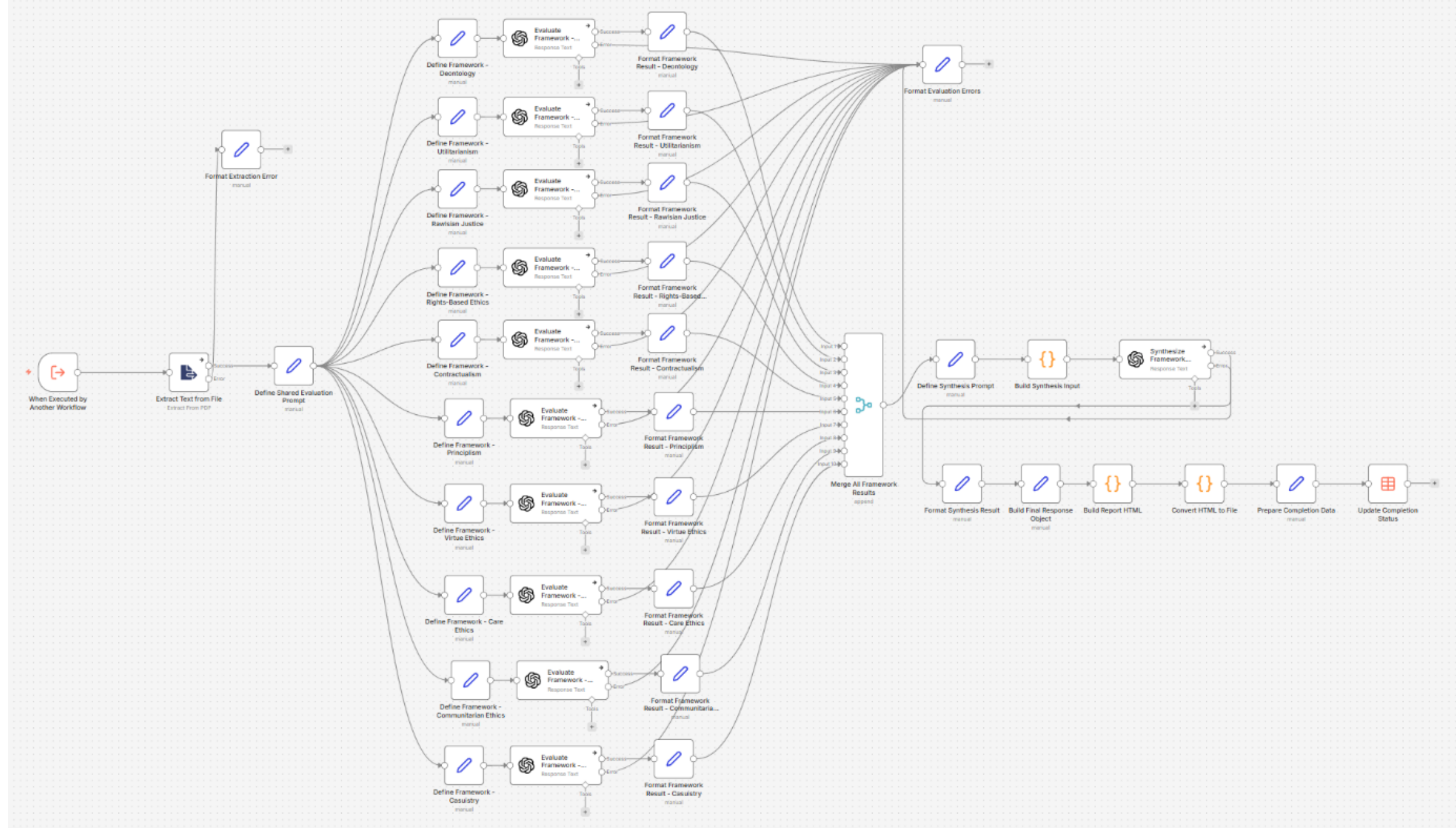

*Figure 3: Worker Workflow*

**Prompt design:** Two prompts were designed: an *evaluator* prompt and a *synthesis* prompt. The evaluator prompt instructed the model to reconstruct the use-case along with the ethical framework (so as to validate fidelity to the supplied documents – use-case and framework description), provide a principle-by-principle critical evaluation, stress-test the use-case against the respective framework, and highlight areas for reflection. The evaluator prompt included explicit non-negotiable constraints (e.g., fidelity to framework and use-case, no fabrications/hallucinations etc.) as well as instructions about the quality, tone, and formatting of output. A benchmarking guide (for inter-framework comparability) was also supplied within the evaluator prompt.

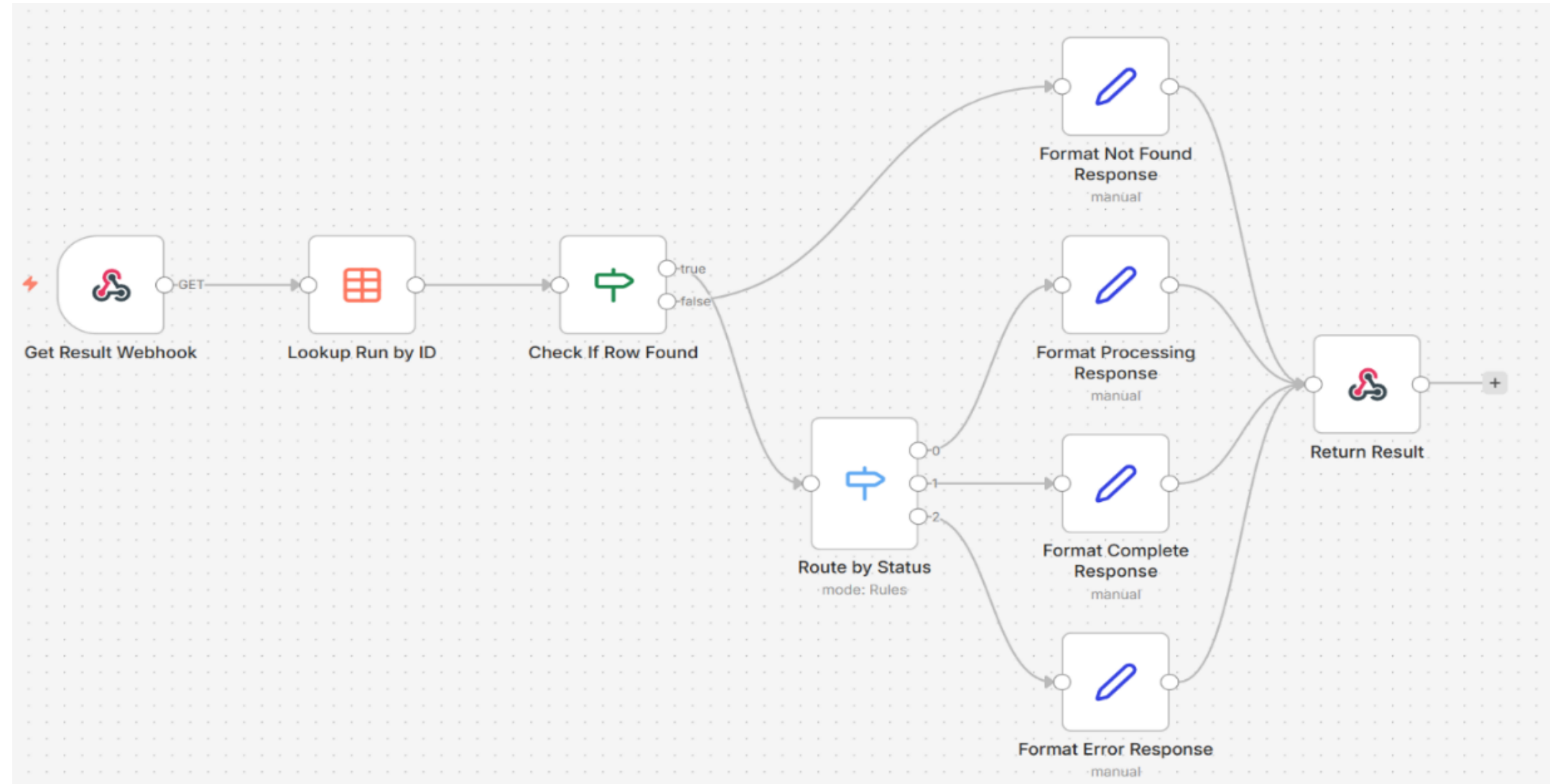

*Figure 2: Emitter Workflow*

The synthesis prompt integrated individual framework evaluations into a single accessible output. It required the LLM to provide a short description of the use-case, its alignment and convergence with different frameworks, how the use-case may be amended to increase cross-framework alignment and to highlight where unresolvable tensions remain. Non-negotiable constraints included fidelity (to the supplied evaluations and use-case), the avoidance of fabrication/hallucinations, and traceability (requiring the synthesizing LLM to explicitly reference the provided evaluations). Tone, quality and accessibility were specified in the prompt to balance philosophical depth with readability for non-specialist users.

**Use-case selection:** Proof-of-concept testing was conducted using three published use-cases from distinct domains. Use-case 1 concerned a non-monetary 'Karma' mechanism for urban congestion management (Riehl et al. 2024), use-case 2 concerned peer-to-peer energy trading with load prioritisation in smart homes (Ahmed et al. 2023), and use-case 3 focused on algorithmic approaches to the allocation of scarce resources within social services (Kube et al. 2023). Use-cases were chosen on the basis that they proposed a prescriptive technical solution to an ethical dilemma.

**Output assessment and analysis:** Generated outputs were analysed qualitatively (without evaluation of the veracity) by the research team across four dimensions: First, that EeVA clearly differentiated between respective frameworks rather than repeating the same conclusion. Second, that outputs presented reasonable ethical depth by identifying and discussing aspects such as stakeholders, framework/use-case assumptions, tensions, recommendations, and unresolved trade-offs. Third, the quality of synthesis was evaluated to identify if the final synthesis integrated framework evaluations into a comparative ethical landscape rather than simply summarising sequentially. Fourth, that outputs were accessible (i.e., readable by a non-specialist), provided clarity, and were of practical use to support further reflection on the ethical implications of respective use-cases.

Recognising the highly interpretive nature of moral philosophy, the fidelity of LLM evaluations against respective framework descriptions was loosely evaluated. Analysis focused on system construction, structured outputs and qualitative assessment rather than on benchmark validation, inter-rater reliability, or formal user evaluation.

# III. RESULTS

## PROTOTYPE IMPLEMENTATION AND OUTPUT GENERATION

In its present proof-of-concept form, EeVA successfully implemented its core objectives. The system accepted a use-case in the form of a file (PDF), generated a unique run identifier, extracted text, processed the use-case through multiple ethical frameworks to output individual framework-specific evaluations, produced a synthesis of those evaluations (including relative metrics), stored the results in a data table, and returned the completed output in a user-friendly format. The separation of upload, processing, and record retrieval ensured stability and usability during longer-running evaluations.

Across three test cases, the system returned consistently structured outputs. Each completed evaluation (≈3000 words) included: a reconstruction of the use-case and ethical framework, a principle-by-principle critical evaluation, a stress test addressing edge cases, and recommendations. In addition, each evaluation included a framework-relative metric with an alignment, concern, and confidence score as well as a brief rationale explaining the system's reasoning behind these scores. The observed output structure closely aligned with the given evaluation prompt and the overall design aim of EeVA.

Each use-case output included a multi-framework synthesis (≈1000 words) that clearly outlined which frameworks strongly aligned or diverged from the use-case, a cross-framework pattern describing ethical issues presented by the use-case that appeared across the different frameworks and a concise overall ethical position summary of the use-case within the broader ethical landscape. A simple graphical representation of the framework-relative metric results was presented, including alignment, concern and confidence levels for each framework evaluation. The synthesis presented a short, structured summary of the 10 framework evaluations, including a brief introduction, an analysis of where different frameworks converge and diverge with the use-case, and an identification of key ethical tensions between the use-case and individual frameworks. The synthesis summarised the recommendations of the evaluation reports as well as where alignment may not be possible.

## QUALITATIVE CHARACTERISTICS OF OUTPUTS

A review of the generated outputs by the authors (one holding a PhD in ethics and another currently pursuing their PhD in ethics) identified four recurring characteristics.

First, the evaluations produced by the evaluation AIs were consistently framework-differentiated. Outputs did not merely restate an ethical judgement, but were directly applied to the use-cases in light of a specific ethical framework. For example, in use-case 1, the evaluation and synthesis AIs showed that several frameworks supported the move away from wealth-based congestion pricing for different reasons (e.g., welfare gain, fairness, reciprocity, or relational responsiveness). At the same time, deontological and contractual perspectives expressed more rigid concerns, particularly regarding urgent exceptions, structural disadvantages, and public legitimacy.

Second, outputs demonstrated ethical depth. They went beyond merely listing advantages and disadvantages to identifying and discussing convergences, fault lines, suggestions for improvements, and highlighting where full ethical alignment between the use-case and specific frameworks was unlikely. For example, in use-case 1, the system pointed to broadly favourable cross-framework agreement of the use-case's attempt to reduce the role of wealth in access to mobility. In addition, EeVA highlighted that ethical support was conditional on the provision of emergency exceptions or protections against structural disadvantages. In use-case 2, tensions between efficiency, inclusion, rights and governance were emphasised in the evaluation, while in use-case 3, the moral difficulty of aggregated gains that leave some households worse off was foregrounded.

Third, EeVA produced a well-developed integrated synthesis. The system not only produced multiple framework essays, but also combined them into a cohesive format accessible to readers without formal training in ethics. For example, in use-case 1, the synthesis explicitly rendered the use-case's proposal favourable across multiple ethical frameworks but highlighted necessary conditions that included the need to consider 'exceptions, institutional embedding, and responsiveness to unequal social conditions.'

Fourth, outputs were readable and practically reflective for non-specialist audiences. They avoided framing conclusions as clearly positive or negative, which may indicate that a use-case's proposal is either ethical or non-ethical. Rather, outputs identified points at which further reflection, redesign, justification or safeguards may be warranted. Consistent with the system's design, the output favoured nuanced guidance over reductive and simplistic responses, directing attention to areas where further reflection may be helpful.

### GENERAL OBSERVATIONS

While the three use-cases differed markedly in their underlying moral structure and the context in which they were applied, EeVA was able to provide ethically coherent responses across the use-cases and ethical frameworks. In each use-case, outputs remained structurally coherent and consistently oriented away from simplistic right or wrong answers toward encouraging ethical reflection. The system identified morally important aims while highlighting unresolved ethical tensions and suggesting areas of reflection to promote cross-framework alignment, while simultaneously recognising that complete cross-framework alignment was unrealistic. This suggests that the prototype functioned as designed: to provide a comparative ethical reflection tool across heterogeneous use-cases.

## IV. DISCUSSION

In the introduction, we highlighted a persistent disconnect between ethically trained professionals and non-ethically trained technical personnel (NTTP) tasked with implementing technical solutions to ethically laden challenges. We laid out the disconnect in terms of a misunderstanding as to the role and function of ethical deliberation. Technical practice often demands operational clarity, whereas ethical reasoning rarely yields univocality. The findings outlined above suggest that LLMs may offer a promising avenue to address this disconnect. The prototype system, EeVA, was explicitly designed to avoid delivering single, ethical verdicts and instead to promote ethical reflection. In doing so, it may represent one step along the way to assisting NTTP in integrating ethical deliberation into technical solutions to ethically laden challenges. EeVA's agentic workflow demonstrated that, in principle, it is possible for LLMs to be organised into a structured system and deployed to provide coherent comparative ethical reflections across heterogeneous socio-technical use-cases.

EeVA was able to operationalise its design aims in a practically usable manner. Applied to three substantively different use-cases – sustainable urban mobility, peer-to-peer distributed energy management, and the allocation of social services in resource-limited contexts – EeVA generated structured outputs that combined framework-specific evaluations with an integrated synthesis. These use-cases were deliberately selected for their diversity, differing not only in their context but also in the normative questions they raised and the kinds of solutions they proposed. The fact that EeVA could be applied across all three use-cases suggests that it is not confined to a single, narrow technical domain but may be adaptable to a wide range of diverse socio-technical challenges.

A central contribution of EeVA lies in its multi-framework structure. Some current attempts to operationalise ethics in technical systems often attempt to reduce ethical reasoning to a limited set of manageable frameworks, checklists, or quantifiable scoring systems. For example, work by the Technical University of Munich (TUM) on motion planning algorithms for autonomous vehicles (AVs) has attempted to algorithmise critical ethical principles such as fairness, equality, and distributed risk (Geisslinger et al. 2021, 2023). While solutions such as these can be operationally attractive, they fail to take into consideration the complexity of ethical action by underplaying genuine normative divergences and reducing moral actions to a single optimisation problem. EeVA points its users in a different direction. Rather than collapsing normative disagreement, it attempts to demonstrate how a single technical proposal can be morally attractive within some frameworks, partially justified under others, but at the same time seriously problematic from other viable ethical frameworks. This is not a weakness of the prototype but an important strength, as it more accurately reflects the structure of ethical reasoning than narrow, simplistic answers to complex questions.

Use-case 1 illustrates this contribution well. The ethical evaluations of the proposed Karma-based solution to urban mobility did not simply label the proposal as fair or unfair. Instead, EeVA demonstrated broad cross-framework support for moving away from wealth-based congestion pricing. This aspect of use-case 1's proposal was generally considered morally promising across multiple ethical frameworks. However, EeVA also pointed to the need for conditions such as making allowances for emergency exceptions, structural disadvantages and public legitimacy to realise this moral promise. EeVA's output, therefore, did not provide a simplistic affirmation or negation of use-case 1's proposal. EeVA drew attention to both the moral promise and the risks in a dynamic, organic way mimicking the ethical deliberation of trained professionals. This is the kind of contribution that may be most valuable to non-ethically trained professionals working in contexts where ethical reflection is often needed but inaccessible; not to deliver final answers, but to identify where further justification, revisions or safeguards are needed.

A further important finding concerns usability. EeVA's outputs were more than abstract philosophical summaries. They were intentionally structured in such a way that a non-ethically trained professional could access and use them to orient their reflection. This is important considering the current practical challenges mentioned in the introduction to this paper. Embedding trained ethicists in technological settings is often impractical. Any system that attempts to improve ethical deliberation must, therefore, be accessible to well-educated, highly skilled technical personnel who are not philosophically trained. Such systems must form a meaningful bridge between philosophical deliberation and operational decision-making. EeVA is designed not to produce ready-to-use ethical answers, but as a way of directing the power of LLMs toward comparative moral reflection so as to provide a scaffold for more ethically informed conversations among non-ethically trained professionals.

The findings presented here should be interpreted cautiously. They do not demonstrate that EeVA can replace trained ethicists with outputs that are normatively prescriptive. Nor do the results demonstrate stable real-world effectiveness in supporting non-ethically trained personnel. Formal user evaluation has yet to be undertaken. What the results do demonstrate is that EeVA is more descriptive yet nevertheless significant: that a prototype agentic AI-driven workflow can be built in a way that preserves ethical plurality, produces structured comparative outputs, and encourages reflective engagement in heterogeneous socio-technical contexts.

Taken together, these results represent a proof-of-concept, not a finished artefact. EeVA does not solve ethical problems in the sense of delivering single *prima facie* authoritative answers. Rather, it helps users locate their proposals within a broader ethical landscape, highlighting questions and areas for reflection that require genuine human judgement. In contexts where access to ethically trained professionals is limited, and technical teams require structured reflection, a system like EeVA may make a meaningful contribution.

## V. LIMITATIONS AND FUTURE DEVELOPMENTS

The study reported here remains a prototype and proof-of-concept, and as such, it is subject to several significant limitations. First, the design and outputs of EeVA exhibit a high degree of non-reproducibility. We can note four reasons why this is the case. A) The probabilistic rather than deterministic nature of LLMs limits reproducibility. This issue is compounded by the use of only OpenAI's ChatGPT 4.1 and 5.4 models. It is likely that other models (e.g., from Gemini or Claude) may have returned different results. B) Framework selection shaped the ethical landscape being evaluated. Other traditions or religious perspectives may have produced different frameworks or used different interpretations of the selected frameworks. C) The selection of use-cases may have influenced the types of outputs produced by EeVA. D) Prompts were developed within the research team and most likely influence outputs. Even the use of English to prompt the tool may have influenced outputs (Agarwal et al. 2024). All this calls into question the reproducibility of EeVA's outputs.

Secondly, human evaluation of EeVA's outputs was limited. While the research team included expert ethicists, no systematic comparison was performed between a human ethicist's output and EeVA's outputs so as to evaluate the veracity of EeVA's evaluations or synthesis. Furthermore, no user testing was conducted as to what end-users' perceptions of EeVA's outputs and usability would be.

Third, costs remain high for the prototype. On average a single execution required 702,666 tokens, took approximately 12:10 minutes and cost $0.46. While acceptable for prototyping, such resource demands are unrealistic to be sustainable at scale without significant efficiency improvements.

To address these limitations, further development and testing are sorely needed. Future iterations should include cross-platform evaluations incorporating multiple LLM providers (e.g., Gemini or Claude), as well as greater diversity in ethical framework descriptions, use-cases, and prompts. Exploration of more advanced AI agentic configurations may also be valuable – for example, allowing an AI agent to take some control of the workflow by deciding which use-cases should be evaluated by which LLMs and which frameworks to use. Crucially, more human involvement (ethicists and end-users) is imperative to understand and improve the usefulness of a tool like EeVA.

## VI. CONCLUSIONS

This paper reports on the development and proof-of-concept testing of a prototype tool designed to assist the ethical reflection of non-ethically trained technical personnel (NTTP). We began by highlighting a significant challenge of communication between ethically-trained experts and NTTP; the expectational mismatch between fields of reference. NTTP require clear, quantifiable solutions to their challenges that may be incorporated into measurable outputs (algorithms, mathematical formulas, programming architecture etc.). Moral reasoning, however, is designed to deal with 'wicked' problems whose very formulations are contested, and whose solutions are not presented as true/false or right/wrong, but rather situated within the context of numerous ethical frameworks. In an ideal world, NTTP would enter into extended discussions with multiple ethicists to reflect on their technical proposals in order to better ensure more ethical outcomes. Limited resources (time, money, personnel etc.) often make this impossible. The advent of LLMs may offer a way to mitigate this challenge. This study produced a working prototype (EeVA) of an agentic-like LLM-based workflow that takes a use-case, evaluates it using 10 different ethical frameworks and produces a readable synthesis with a focus on promoting ethical reflection.

The results presented here of the proof-of-concept testing are promising. Across three use-cases in distinct contexts, EeVA was able to produce ethically coherent evaluations across all 10 frameworks as well as accessible syntheses that included areas of convergence, divergence, recommendations to increase cross-framework convergence as well as an indication of where irreconcilable differences remain. EeVA's outputs were consistently well-structured, aligned with the underlying framework analyses, and demonstrated a level of interpretive insight that suggests potential value as a support tool for NTTP engaging in ethical deliberation.

While our discussions here have focused on technology and technical solutions to ethical problems, EeVA is not limited to this context. There is no reason the logic underpinning EeVA would not be able to perform a similar function in non-technical contexts such as social care, public policy, healthcare, etc. Indeed, wherever a solution is needed to an ethically laden problem, and access to trained ethicists is limited, EeVA may be useful.

Before concluding, it is important to emphasise that EeVA is subject to numerous limitations that include the nature of its dynamic, non-reproducible outputs, its lack of extensive human testing (both from trained ethicists and end-users), and running costs (time, tokens, financial etc.). While the limitations are serious challenges, they do not negate the purpose and function of EeVA as a prototype and proof-of-concept. Ethics is, by its very nature, dynamic and non-reproducible. This project did not aim to design an AI tool that could produce benchmarked objective ethical answers to complex technical questions. Rather, the tool is designed to provide a scaffold for ethical reflections by NTTP. In this context, dynamic, changing outputs may be viewed as a strength rather than a limitation. Nevertheless, it must be stressed that EeVA should not be used as the basis of sound ethical reasoning, nor deployed as a substitute for human moral philosophising.

## VII. FUNDING STATEMENT

The research leading to these results has received funding from the Horizon Europe Programme under the Marie Skłodowska-Curie grant agreement No 101183031. Views and opinions expressed are however those of the author(s) only and do not necessarily reflect those of the European Union. Neither the European Union nor the granting authority can be held responsible for them. In addition, the research has been supported by the NCCR (Automation), a National Centre of Competence (or Excellence) in Research, funded by the Swiss National Science Foundation (Grant Number 51NF40_225155).

## VIII. DATA AVAILABILITY

The materials underlying this study, including the EeVA workflow and associated files are available from the corresponding author on reasonable request.